\documentclass[aps,showpacs,showkeys,superscriptaddress]{revtex4}

\usepackage{graphicx}
\usepackage{makeidx}
\usepackage[]{hyperref}
\usepackage{amsmath}

\usepackage{amssymb,epsfig,graphics}
\usepackage{amscd}
\usepackage{verbatim}

\usepackage{amsmath}
\usepackage{amsthm}
\usepackage{amsfonts}

\usepackage{tabularx}

\begin{document}

\newcolumntype{L}[1]{>{\raggedright\arraybackslash}p{#1}}
\newcolumntype{C}[1]{>{\centering\arraybackslash}p{#1}}
\newcolumntype{R}[1]{>{\raggedleft\arraybackslash}p{#1}}

\title{Electronic Properties of Bilayer Fullerene Onions}

\author{R. Pincak}\email{pincak@saske.sk}
\affiliation{Institute of Experimental Physics, Slovak Academy of Sciences,
Watsonova 47,043 53 Kosice, Slovak Republic}
\affiliation{Bogoliubov Laboratory of Theoretical Physics, Joint
Institute for Nuclear Research, 141980 Dubna, Moscow region, Russia}

\author{V. V. Shunaev}\email{vshunaev@list.ru}
\affiliation{Saratov State University, Astrakhanskaya Street 83, 410012 Saratov, Russia}

\author{J. Smotlacha}\email{smota@centrum.cz}
\affiliation{Bogoliubov Laboratory of Theoretical Physics, Joint
Institute for Nuclear Research, 141980 Dubna, Moscow region, Russia}

\author{M. M. Slepchenkov}\email{slepchenkovm@mail.ru}
\affiliation{Saratov State University, Astrakhanskaya Street 83, 410012 Saratov, Russia}

\author{O. E. Glukhova}\email{oeglukhova@yandex.ru}
\affiliation{Saratov State University, Astrakhanskaya Street 83, 410012 Saratov, Russia}

\begin{abstract}
The HOMO-LUMO gaps of the bilayer fullerene onions were investigated. For this purpose, the HOMO and LUMO energies were calculated for the isolated fullerenes using the parametrization of the tight binding method with the Harrison-Goodwin modification. Next, the difference of the Fermi levels of the outer and inner shell was calculated by considering the hybridization of the orbitals on the base of the geometric parameters. The results were obtained by the combination of these calculations.

\end{abstract}
\date{\today}




\maketitle

\section{Introduction}

Carbon onions (or fullerene onions) are concentric fullerenes nested in each other. For the first time their formation was observed in 1992 by Ugarte who focused an electron beam in a sample of amorphous carbon \cite{1}. Nowadays carbon onions can be produced by nanodiamond annealing \cite{2}, arc discharge between two graphite electrodes in water \cite{3,4} or by the naphthalene combustion \cite{5}. Unique properties of carbon onions make them the element base of various electric devices. Carbon onions are used as components for the electric double layer capacitors, also called supercapacitors \cite{6,7}. Pech et al. prepared ultrahigh-power micrometer-sized supercapacitors based on the carbon onions \cite{8}. In combination with ${\rm Co_3O_4}$ and ${\rm MnO_2}$ carbon onions can be used as electrode materials for ion-lithium batteries \cite{9,10}. The nanofilters based on the  carbon onions and its composites can be applied as electromagnetic interference shields for the terahertz waves \cite{11}. Application of the carbon onions in electric devices requires the information about its electron parameters like energy gap and Fermi levels.

In the case of the bilayer onions which we investigate in this paper, the energy gap is given by the HOMO-LUMO gap - the energy difference between the lowest unoccupied molecular orbital of the inner onion shell and of the highest occupied molecular orbital of the outer onion shell. This difference is influenced by the energy difference between the Fermi levels of both onion shells.

\begin{figure}[htbp]
\includegraphics[width=6cm]{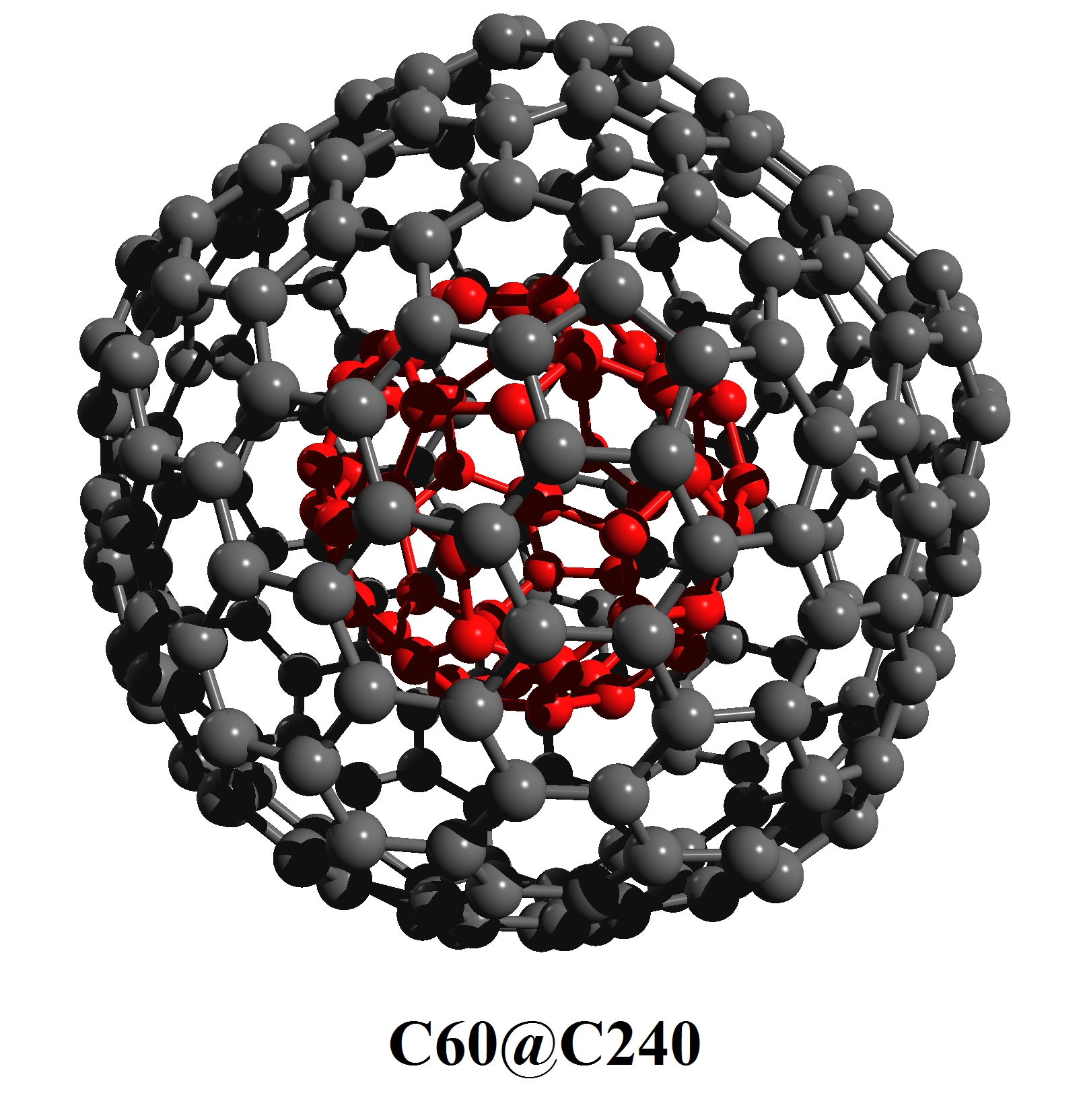}
\caption{Bilayer carbon onion.}\label{fg1}
\end{figure}

In our previous paper we calculated the energy gap of the onion ${\rm C_{60}@C_{240}}$ (Fig. \ref{fg1}) by the treatment of the hybridization of the fullerene orbitals \cite{12}. To calculate this value we took the experimental data about isolated fullerenes ${\rm C_{60}}$ and ${\rm C_{240}}$ \cite{13,14}. To expand our knowledge about the electronic properties of other bilayer onions we need information about their isolated part. In this paper we will obtain this information due to tight-binding method that was successfully used for the investigation of the electronic and the geometry properties of the bilayer onions in \cite{15}.

In section 2, we describe the tight-binding model with the original parametrization. On the base of this model we receive the geometry parameters which are necessary for the calculation of the Fermi levels of the considered onions. We also calculate the van der Waals interaction and on this base we determine which forms of the carbon onions can exist.
In section 3, we briefly describe how to obtain the HOMO-LUMO gaps for the bilayer onions by the combination of the calculations of the Fermi levels and of the HOMO and LUMO energies. It is described in \cite{12} how to calculate the energy difference between the Fermi levels. They are calculated due to parametrization based on the hybridization of the orbitals. The geometry parameters received in previous section are used here.\\

\section{The results obtained by tight-binding model with original parametrization}

To calculate the geometry and the electronic structure of the isolated fullerenes, the original parametrization of the tight binding method with the Harrison-Goodwin modification was used. To obtain the metric features of the investigated object, the total energy $E$ is minimized on the bond lengths:

\begin{equation}E = {E_{bond}} + {E_{rep}},\end{equation}
where $E_{bond}$ is the energy of occupied energy states (levels), $E_{rep}$ is the repulsive energy with an account of internuclear and electron-electron interaction. The energy of the band structure is determined by the formula

\begin{equation}{E_{bond}} = 2\mathop \sum \limits_n {\varepsilon _n},\end{equation}
where $\varepsilon_n$ is $n-$th energy level that corresponds to an eigenvalue of the Hamiltonian (multiplier 2 takes into account spin of electron). The repulsion energy is represented as the sum of pair potentials:

\begin{equation}{E_{rep}} = \mathop \sum \limits_{i < j} {V_{rep}}\left( {\left| {{r_i} - {r_j}} \right|} \right)\end{equation}
where $i, j$ - the numbers of interacting atoms, $r_i, r_j$ - Cartesian coordinates. The term $V_{rep}$ is determined by the Goodwin's expression \cite{16}:

\begin{equation}{V_{rep}}\left( r \right) = V_{rep}^0{\left(\frac{{1.54}}{r}\right)^{4.455}} \times\exp\left\{4.455\left[ - {{\left( {\frac{r}{{2.32}}} \right)}^{22}} + {{\left( {\frac{{1.54}}{{2.32}}} \right)}^{22}}\right] \right\},\end{equation}
where $V_{rep}^0= 10.92$ eV.

To find $E_{bond}$ we need to fill in the Hamiltonian. For this case we should determine carbon terms $\varepsilon_s$ and $\varepsilon_p$, and the overlapping integrals $V_{ss\sigma }^0,V_{sp\sigma }^0,V_{pp\sigma }^0,V_{pp\pi }^0$. The interatomic matrix elements of Hamiltonian were determined as follows \cite{19}:

\begin{equation}{V_{ij\alpha }}\left( r \right) = V_{ij\alpha }^0{\left(\frac{{1.54}}{r}\right)^{2.796}} \times \exp\left\{ {2.796\left[ - {{\left( {\frac{r}{{2.32}}} \right)}^{22}} + {{\left( {\frac{{1.54}}{{2.32}}} \right)}^{22}}\right]} \right\},\end{equation}
where $r$ is the distance between the atoms, $i, j$ - orbital moments of the wave functions, $\alpha$ is the index noting a bond type ($\sigma$ or $\pi$). The initial Goodwin's modification had the array of disadvantages: it didn't allow to calculate the potential of ionization and the energy gaps obtained by this method differed from the experimental data. So we developed the original parametrization of the Hamiltonian by the comparison of the experimental data for fullerene ${\rm C_{60}}$ (bond length $r_1$ and $r_2$, energy gap $E_g$ and ionization potential $I$ \cite{17}) with the calculated. The obtained matrix elements of the Hamiltonian are shown in Table \ref{tab1}. Herewith for the fullerene ${\rm C_{60}}$ it was found: ${r_1} = 1.4495\,{\rm \AA},\,\,{r_2} = 1.4005\,{\rm \AA},\,\,{E_g} = 1.96\,{\rm eV},\,\,I = 7.6099\,{\rm eV}$.

\renewcommand{\arraystretch}{2}

\begin{table}[htbp]

\caption{Atom terms of carbon and ground overlapping integrals (in eV).}
\begin{tabular}{|C{1.5cm}||C{1.5cm}|C{1.5cm}|C{1.5cm}|C{1.5cm}|C{1.5cm}|C{1.5cm}|}
\hline
term & $\varepsilon_s$ & $\varepsilon_p$ & $V_{ss\sigma}^0$ & $V_{sp\sigma}^0$ & $V_{pp\sigma}^0$ & $V_{pp\pi}^0$\\
\hline
value & -10.932 & -5.991 & -4.344 & 3.969 & 5.457 & -1.938\\
\hline
\end{tabular}

\label{tab1}

\end{table}

\renewcommand{\arraystretch}{1}

Let's note that our parametrization of the Harrison-Goodwin tight-binding modification allows to calculate the energy gap and the potential of ionization of isolated fullerenes with good accuracy, though it can't find the electron structure of the bilayer onions. But it can be used to find the ground state of them and, thus, its geometry parameters. We have already successfully applied this method to find the geometry structure and the ground states of the molecules ${\rm C_{20}@C_{240}}$ and ${\rm C_{60}@C_{540}}$ \cite{15}. In this paper we expanded the number of considered isolated fullerenes and onions. The HOMO and LUMO gaps of the isolated fullerenes are in Table \ref{tab1a} and their geometry parameters are given in Table \ref{tab2}.

\renewcommand{\arraystretch}{2}

\begin{table}

\caption{Energy in eV of HOMO and LUMO energies and the corresponding gaps for isolated fullerenes.}
\begin{tabular}{|C{1.7cm}||C{1.7cm}|C{1.7cm}|C{1.7cm}|C{1.7cm}|C{1.7cm}|C{1.7cm}|C{1.7cm}|C{1.7cm}|}
\hline
& ${\rm C_{20}}$ & ${\rm C_{28}}$ & ${\rm C_{32}}$ & ${\rm C_{36}}$ & ${\rm C_{60}}$ & ${\rm C_{80}}$ & ${\rm C_{240}}$ & ${\rm C_{540}}$\\ \hline\hline
HOMO & $-6.7$ & $-7.19$ & $-6.85$ & $-7.02$ & $-7.6$ & $-6.81$ & $-7.13$ & $-7.02$\\ \hline
LUMO & $-6.31$ & $-7.02$ & $-5.8$ & $-6.92$ & $-5.57$ & $-6.7$ & $-5.88$ & $-5.93$\\ \hline
H-L gap & $0.38$ & $0.17$ & $1.04$ & $0.09$ & $2.03$ & $0.1$ & $1.25$ & $1.09$\\ \hline
\end{tabular}

\label{tab1a}

\end{table}

\renewcommand{\arraystretch}{1}

\begin{table}[htbp]

\caption{The geometry parameters of some isolated fullerene obtained by original tight-binding method.}
\begin{tabular}{|C{3cm}|C{3cm}|C{3cm}|}
\hline
Fullerene & Radius,${\rm \,\AA}$ & Average bond length,${\rm \,\AA}$\\
\hline\hline
${\rm C_{20}}$ & 2.06 & 1.47\\ \hline
${\rm C_{28}}$ & 2.53 & 1.46\\ \hline
${\rm C_{32}}$ & 2.65 & 1.51\\ \hline
${\rm C_{36}}$ & 2.37 & 1.44\\ \hline
${\rm C_{60}}$ & 3.4 &	1.445\\ \hline
${\rm C_{80}}$ & 3.94 & 1.44\\ \hline
${\rm C_{240}}$ & 7.2 & 1.415\\ \hline
${\rm C_{540}}$ & 10.1 & 1.435\\ 
\hline
\end{tabular}

\label{tab2}

\end{table}

The intershell interaction in bilayer onions is mainly determined by the van der Waals interaction and the overlapping energy of the electron clouds. We used the potential of Lennard-Jones to determine the interaction between the non-bonded atoms. As we showed earlier \cite{15}, for the bilayer onions with external icosahedral shell (${\rm C_{60}}$, ${\rm C_{240}}$, ${\rm C_{540}}$) there are three potential wells where the internal fullerene can be located. In this paper, for such onions we will consider the well with the lowest energy. The intershell interaction between the internal and the external shells of the onions is shown in Table \ref{tab3}.

\begin{table}

\caption{The intershell interaction (van der Waals interaction) for some onions (in eV, model \#1).}
\begin{tabular}{|C{2cm}||C{2cm}|C{2cm}|C{2cm}|C{2cm}|C{2cm}|C{2cm}|C{2cm}|}
\hline
${\rm C_{80}}$ & -1.256 & X & X & X & X & X & X \\ \hline
${\rm C_{240}}$ & -0.786 & -1.066 &	-1.228 & -1.228 & -2.66 & -2.41 & X \\ \hline
${\rm C_{540}}$ & -0.518 & -0.656 &	-0.658 & -0.743 & -1.209 & -1.16 & -7.106 \\ \hline\hline
@ &	${\rm C_{20}}$ & ${\rm C_{28}}$ & ${\rm C_{32}}$ & ${\rm C_{36}}$ &	${\rm C_{60}}$ & ${\rm C_{80}}$ &
${\rm C_{240}}$ \\ \hline
\end{tabular}

\label{tab3}

\end{table}

Let's note that the energy of the van der Waals interaction is positive for the onions with such external shells as ${\rm C_{20}}$, ${\rm C_{28}}$, ${\rm C_{32}}$, ${\rm C_{36}}$, ${\rm C_{60}}$. It means that bilayer onions with these external shells can't exist under usual conditions. The onion ${\rm C_{20}@C_{80}}$ is the only bilayer onion with external shell ${\rm C_{80}}$ that has negative energy.\\

\section{The results obtained by the parametrization based on the hybridization of the orbitals}
The main content of this method consists in the calculation of the difference between the Fermi levels of the outer and the inner shell, given by the energy of the $\pi$-bonds which are perpendicular to the molecular surface. In \cite{12}, the spatial orientation of the corresponding bonds is considered for this purpose. Here, the wave function of the $\pi$-bond corresponding to the inner sphere of the bilayer onion has the form
\begin{equation}|\pi \rangle  = {D_1}|s\rangle  + {D_2}|{p_x}\rangle  + {D_4}|{p_z}\rangle,\end{equation}
where the following equations are satisfied:
\begin{equation}\langle {\sigma _i}|{\sigma _j}\rangle  = {\delta _{ij}},\langle \pi |{\sigma _j}\rangle  = 0,\langle \pi |\pi \rangle  = 1.\end{equation}
Here, $|\sigma_i\rangle$  correspond to the bonds lying on the molecular surface. The values of $D_1, D_2, D_4$   depend on the radius and average bond length of the given form of fullerene. For different kinds of fullerenes, these geometry parameters were calculated in the previous section.

Then, the energy of the corresponding $\pi$-bond is given by
\begin{equation}\epsilon = \langle \pi |H|\pi \rangle  = D_1^2\langle s|H|s\rangle  + D_2^2\langle {p_x}|H|{p_x}\rangle  + D_4^2\langle {p_z}|H|{p_z}\rangle, \end{equation}
where \cite{18}
\begin{equation}\langle s|H|s\rangle  \approx  - 12{\rm{eV}},\langle {p_x}|H|{p_x}\rangle {\rm{ = }}\langle {p_y}|H|{p_y}\rangle {\rm{ = }}\langle {p_z}|H|{p_z}\rangle  \approx  - 4{\rm{eV}}\end{equation}
In Table \ref{tab4}, the values of ${D_1},{D_2},{D_4}$ are listed together with the energy of $\pi$-bonds.

\begin{table}

\caption{Parameters ${D_1},{D_2},{D_4}$ and the energy of $\pi$-bonds (in eV).}
\begin{tabular}{|C{3cm}||C{3cm}|C{3cm}|C{3cm}|C{3cm}|}
\hline
Fullerene & $D_1$ & $D_2$ & $D_4$ & $\varepsilon$ \\ \hline\hline
${\rm C_{20}}$ & 0.225i & -0.517 & 0.885 & -3.595 \\ \hline
${\rm C_{28}}$ & 0.367 & -0.178 & 0.913 & -5.076 \\ \hline
${\rm C_{32}}$ & 0.366 & -0.169	& 0.915	& -5.071 \\ \hline
${\rm C_{36}}$ & 0.364 & -0.223	& 0.904	& -5.060 \\ \hline
${\rm C_{60}}$ & 0.297 & -0.558	& 0.953	& -4.705 \\ \hline
${\rm C_{80}}$ & 0.257 & -0.033	& 0.966	& -4.053 \\ \hline
${\rm C_{240}}$ & 0.139 & -0.005 & 0.990 & -4.155 \\ \hline
${\rm C_{540}}$ & 0.101	& $\sim$ 0 & 0.990	& -4.081 \\ \hline
\end{tabular}

\label{tab4}

\end{table}

On the base of these results, the difference between the Fermi levels of the onion shells was calculated using the formula $\Delta  = {\varepsilon _{out}} - {\varepsilon _{in}}$ with the results listed in Table \ref{tab5}. Regarding the results of the calculation of the corresponding van der Waals interaction (Table \ref{tab3}), we exclude all the unacceptable forms of the onions.

\begin{table}

\caption{Difference between Fermi levels of onion shells (model \#2).}
\begin{tabular}{|C{2cm}||C{2cm}|C{2cm}|C{2cm}|C{2cm}|C{2cm}|C{2cm}|C{2cm}|}
\hline
${\rm C_{80}}$ & -0.936 & X & X & X & X & X & X \\ \hline
${\rm C_{240}}$ & -0.560 & 0.921 & 0.916 & 0.905 & 0.550 & 0.375 & X \\ \hline
${\rm C_{540}}$ & -0.486 & 0.995 & 0.990 & 0.979 & 0.624 & 0.449 & 0.074 \\ \hline\hline
@ &	${\rm C_{20}}$ & ${\rm C_{28}}$ & ${\rm C_{32}}$ & ${\rm C_{36}}$ &	${\rm C_{60}}$ & ${\rm C_{80}}$ &
${\rm C_{240}}$ \\ \hline
\end{tabular}

\label{tab5}

\end{table}

\begin{figure}
\includegraphics[width=15cm]{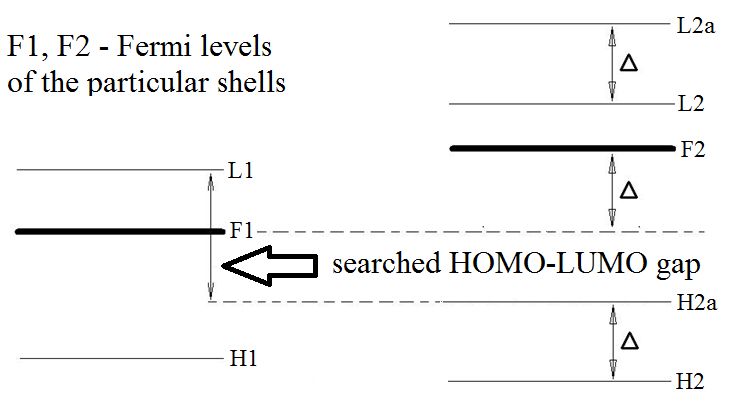}
\caption{Calculation of the HOMO-LUMO gap in bilayer onions.}\label{fg2}
\end{figure}

Now, we can calculate the HOMO-LUMO gap for the bilayer onions. The procedure is given by the scheme in Fig. \ref{fg2}. Here, the symbols $F_1$, $F_2$ correspond to ${\varepsilon _{in}},{\varepsilon _{out}}$ in the above formula - they denote the Fermi levels of the corresponding shells. $H_1$, $L_1$, $H_2$, $L_2$ denote the highest occupied and the lowest unoccupied molecular orbital energy levels of the inner and outer shell, respectively, $H_{2a}$ and $L_{2a}$ denote the $H_2$ and $L_2$ levels shifted by the energy difference $\Delta$ of the Fermi levels. Then, the HOMO-LUMO gap of the bilayer onion can be calculated as
\begin{equation}{E_{H - Lgap}} = {L_1} - {H_{2a}} = {L_1} - ({H_2} + \Delta ) = {L_1} - {H_2} - \Delta \end{equation}

The values of HOMO and LUMO energy levels of isolated fullerenes are calculated using the method described in the previous section. They corresponding HOMO-LUMO gaps are listed in Table \ref{tab1a}.

%
%
%
%
%

Then, using the values of the energy difference between the Fermi levels listed in Table \ref{tab5}, we get the results in Table \ref{tab7}.

\begin{table}

\caption{HOMO-LUMO gap of bilayer onions.}
\begin{tabular}{|C{2cm}||C{2cm}|C{2cm}|C{2cm}|C{2cm}|C{2cm}|C{2cm}|C{2cm}|}
\hline
${\rm C_{80}}$ & 1.44 & X & X & X & X & X & X \\ \hline
${\rm C_{240}}$ & 1.38 & -0.81 & 0.41 &	-0.70 &	1.01 & 0.05 & X \\ \hline
${\rm C_{540}}$ & 1.01 & -1.18 & 0.04 &	-1.07 &	0.64 & -0.32 & 0.88 \\ \hline\hline
@ &	${\rm C_{20}}$ & ${\rm C_{28}}$ & ${\rm C_{32}}$ & ${\rm C_{36}}$ &	${\rm C_{60}}$ & ${\rm C_{80}}$ &
${\rm C_{240}}$ \\ \hline
\end{tabular}

\label{tab7}

\end{table}

\section{Conclusion}

\begin{figure}[htbp]
\includegraphics[width=15cm]{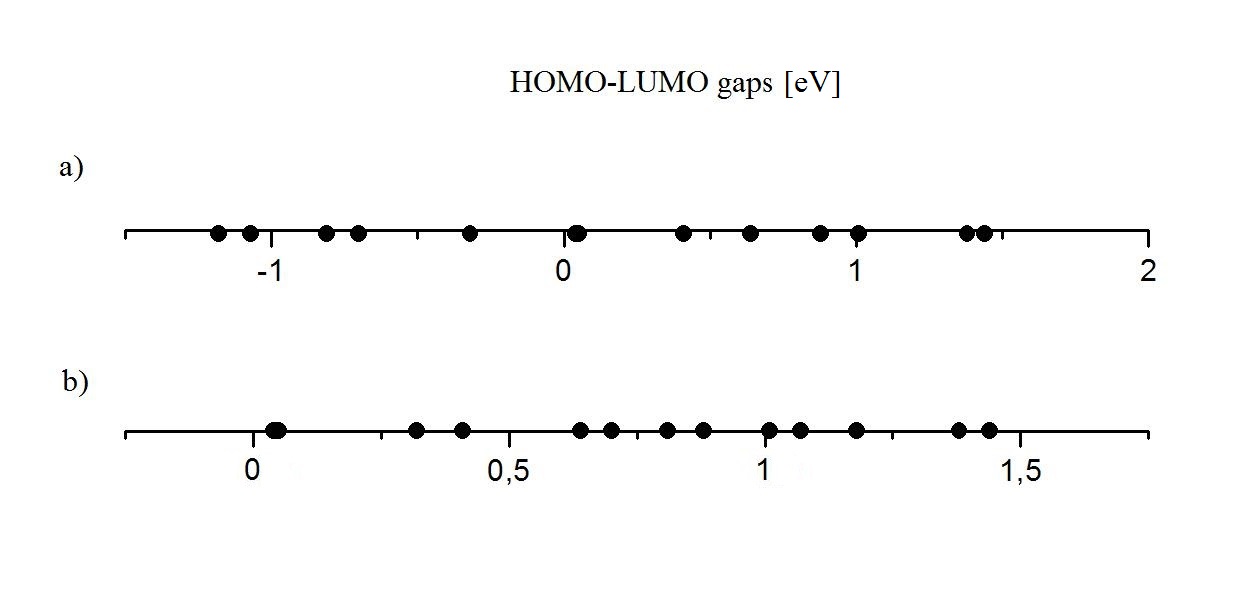}
\caption{Distribution of the HOMO-LUMO gaps on the base of Table \ref{tab7}: a) regarding the order of the energy values for the outer and the inner shell, b) disregarding the order of the energy values for the outer and the inner shell.}\label{fg3}
\end{figure}

We verified which forms of bilayer fullerene onions can really exist and we found the HOMO-LUMO gaps of them which are listed in Table \ref{tab7}. Some of the values have negative sign which means that the value of the LUMO energy level of the inner shell is lower than the HOMO energy level of the outer shell. To get an insight into these results, on the base of Table \ref{tab7}, we can outline their distribution, where we can regard the real values or the modulus (Fig. \ref{fg3}).

From the the outlined distributions does not follow any special rule, for different kinds of the bilayer onions the mutual placement is accidental. It is worth mentioning the coincidence of the HOMO-LUMO gaps for ${\rm C_{20}@C_{540}}$ and ${\rm C_{60}@C_{240}}$ (1.01 eV) and for ${\rm C_{32}@C_{540}}$ and ${\rm C_{80}@C_{240}}$ (0.04 eV and 0.05 eV).

To conclude, the HOMO-LUMO gap serve as a characteristic of the electronic properties: the lower value of this gap, the more metallic the corresponding material is. From our investigations follows that the less metallic is ${\rm C_{20}@C_{80}}$, the most metallic are ${\rm C_{32}@C_{540}}$ and ${\rm C_{80}@C_{240}}$.\\

ACKNOWLEDGEMENTS --- The authors gratefully acknowledge funding of this work by Presidential scholarship 2016-2018 (project No. SP-3135.2016.1).

\end{document}